\documentclass[iop]{emulateapj}

\usepackage{color}
\usepackage{ulem}

\newcommand{\Ks}{K_{\mathrm{s}}}
\newcommand{\JHK}{JH\Ks}
\newcommand{\AK}{A_{\Ks}}

\newcommand{\AKEHK}{A_{\Ks}/E_{H-\Ks}}
\newcommand{\AKEKT}{A_{\Ks}/E_{\Ks - [3.6]}}
\newcommand{\RGC}{R_{\mathrm{GC}}}
\newcommand{\kms}{{\mathrm{km~s}^{-1}}}
\newcommand{\Vhelio}{V_{\mathrm{helio}}}
\newcommand{\VLSR}{V_{\mathrm{LSR}}}

\shorttitle{Classical Cepheids in the inner Galactic disk}
\shortauthors{Tanioka et al.}

\begin{document}

\title{New classical Cepheids in the inner part of\\
the northern Galactic disk and their kinematics}

\received{2017 February 1}
\revised{2017 April 26}
\accepted{2017 May 6}

\author{
Satoshi Tanioka\altaffilmark{1,2},
Noriyuki Matsunaga\altaffilmark{2},
Kei Fukue\altaffilmark{3,4},
Laura Inno\altaffilmark{5},
Giuseppe Bono\altaffilmark{6,7},
and
Naoto Kobayashi\altaffilmark{8,9,4}
}
\altaffiltext{1}{National Astronomical Observatory of Japan, 2-21-1 Osawa, Mitaka, Tokyo 181-8588, Japan}
\altaffiltext{2}{Department of Astronomy, School of Science, The University of Tokyo, 7-3-1 Hongo, Bunkyo-ku, Tokyo 113-0033, Japan; matsunaga@astron.s.u-tokyo.ac.jp}
\altaffiltext{3}{Koyama Astronomical Observatory, Kyoto Sangyo University, Motoyama, Kamigamo, Kita-ku, Kyoto 603-8555, Japan}
\altaffiltext{4}{Laboratory of Infrared High-resolution spectroscopy (LiH), Koyama Astronomical Observatory, Kyoto Sangyo University, Motoyama, Kamigamo, Kita-ku, Kyoto-603-8555, Japan}
\altaffiltext{5}{Max Planck Institute for Astronomy, K\"{o}nigstuhl 17, D-69117 Heidelberg, Germany}
\altaffiltext{6}{Dipartimento di Fisica, Universit\'{a} di Roma Tor Vergata, Via della Ricerca Scientifica 1, 00133 Rome, Italy}
\altaffiltext{7}{Instituto Nazionale de Astrofisica, Osservatorio Astronomico di Roma, Via Frascati 33, 00040 Monte Porzio Catone, Italy}
\altaffiltext{8}{Institute of Astronomy, School of Science, The University of Tokyo, 2-21-1 Osawa, Mitaka, Tokyo 181-0015, Japan}
\altaffiltext{9}{Kiso Observatory, Institute of Astronomy, School of Science, The University of Tokyo, 10762-30 Mitake, Kiso-machi, Kiso-gun, Nagano 397-0101, Japan}

\begin{abstract}
The characteristics of the inner Galaxy remain obscured by significant dust
extinction, and hence infrared surveys are useful to find young Cepheids
whose distances and ages can be accurately determined.
A near-infrared photometric and spectroscopic survey was carried out
and three classical Cepheids were unveiled in the inner disk,
around $20\degr$ and $30\degr$ in Galactic longitude. The targets feature small
Galactocentric distances, 3--5~kpc, and their velocities are important as
they may be under the environmental influence of the Galactic bar.
While one of the Cepheids has radial velocity consistent with
the Galactic rotation, the other two are moving significantly slower. 
We also compare their kinematics with that of
high-mass star-forming regions with parallactic distances measured.
\end{abstract}


\section{Introduction}

The inner part of the Galaxy is a crossroad of various stellar populations,
and different components of the Galaxy like the bulge and the disk are mixed.
Within ${\sim}250$~pc around the supermassive blackhole, Sgr~A$^*$,
stars with a wide range of age and interstellar gas form a disk known as
the nuclear stellar disk or the central molecular zone \citep{Launhardt-2002}.
This relatively small component of the Galaxy (${\sim}10^9~M_\odot$ in mass)
is surrounded by the bulge that is more massive,
${\sim}2\times 10^{10}~M_\odot$ \citep{Valenti-2016}, and more extended.
Further out, extended is the disk of stars and interstellar matter,
the main component of the Galaxy
(${\sim}10^{12}~M_\odot$, \citealt{Bland-Hawthorn-2016}).
In the rest of the Introduction, we give a review on stellar populations
present in the components of the inner Galaxy
and describe the goal of this study in which we use Cepheids
as tracers of stellar kinematics at the innermost part of the disk.

A trouble in studies of the inner Galaxy is interstellar extinction.
It is therefore natural that the earliest studies on stellar populations
of the bulge were focused on stars found in the low-extinction regions 
represented by the Baade window and the outer parts at higher
Galactic latitudes \citep[see the review by][]{Rich-2013}.
The development of infrared instruments enlarged the window 
to study the bulge and other regions in the inner Galaxy
thanks to the significantly reduced effect of interstellar extinction.
For example, in the early 1990s, the bulge was found to be elongated
showing a bar-like structure \citep[e.g.][]{Nakada-1991}
together with the elongated distribution of interstellar gas 
observed in the radio regime \citep{Binney-1991}.

Besides the main bar-like bulge which dominates
the surface brightness distribution at $|l|<10\degr$,
some additional structures have
been suggested to explain the distribution of red clump giants
in the inner part, $|l|<4\degr$\citep[e.g.][]{Nishiyama-2005},
and in the outer part, $10\degr < l < 30\degr$
\citep[e.g.][]{Hammersley-1994,Hammersley-2000}.
It is, however, unclear if there are more than one bars or the structural
parameters of the bar(s) show a smooth variation
when traced by stellar populations with different ages \citep{Wegg-2015}.
In any case, the presence of the extended bar-like distribution of
relatively young populations represented by red clump giants
is an interesting feature of the inner Galaxy, and it may be
related to the presence of even younger stellar clusters, from
a few to ${\sim}50$~Myr, that seem to be preferentially located
near the end of the extended bar \citep{Davies-2009,Davies-2012}.
With the Galactocentric distances of 3--5~kpc, many of these clusters
are located at around the interface of the bar/bulge and the disk.

As a matter of fact, the interface between the Galactic bulge and
the innermost part of the disk has not been explored well compared to
the bulge itself. The bulge has a steep radial profile which falls
down to a density lower than the disk component at
around a couple of kilo-parsecs \citep[e.g.][]{Sofue-2013},
while the Galactic disk is suggested to have
a hole in the central region \citep{Einasto-1979,Robin-2003}.
Radio observations of \ion{H}{2} regions and high-mass star forming regions
(HMSFRs) with maser emission seem to support such a hole by finding
the drop in density of these objects toward the central region of the disk
\citep{Anderson-2011,Anderson-2012,Jones-2013,Sanna-2014}.
Identifying stellar populations in such a central region of the disk is,
however, challenging with several observational difficulties.
For example, various stellar populations are overlaid along the line of sight.
Distances can be determined only for a limited kinds of objects.
Above all, interstellar extinction makes it hard to detect objects
in the optical and complicates interpretations even if they are visible
at longer wavelengths.
 
The goal of this study is to find Cepheids located 
at the interface of the bar/bulge and the disk, and to
study their kinematics.
Cepheids follow the famous period--luminosity relation (PLR, hereinafter) which
works as a fundamental step of the cosmic distance scale
\citep[see e.g.][]{Freedman-2010} and the stars also adhere to
a period--age relation \citep{Bono-2005}. They are therefore useful
as a tracer of young stellar populations (10--300~Myr)
in dust-obscured regions of the Galaxy. However, surveys of distant Cepheids
are incomplete because of extreme extinction toward the Galactic disk.
Recent developments of near-infrared observing facilities and their
systematic surveys have opened the new path to such obscured Cepheids
as demonstrated by recent works
\citep{Matsunaga-2011b,Feast-2014,Dekany-2015a,Dekany-2015b,Matsunaga-2016}.
In particular, \citet{Matsunaga-2016} reported that the density of Cepheids
located within ${\sim}2.5$~kpc from the Galactic center
is lower than the surrounding part except the nuclear stellar disk,
which is consistent with the aforementioned radio observations.

In this paper, we report results of our new
survey for small regions toward the inner disk, $20\degr$ and $30\degr$
in Galactic longitude, and spectroscopic follow-up observations
for Cepheids discovered.
While the new Cepheids are located in a different region of
the disk compared to those previously found in the series of
our near-infrared searches for Cepheids \citep{Matsunaga-2011b,Matsunaga-2016}, 
an important step in this paper is
to demonstrate a method of comparing the radial velocities of Cepheids
with the Galactic rotation with relevant uncertainties,
{e.g.}~in distance, taken into account.

\section{Photometric observations}

\subsection{Data and analysis}

We used the IRSF telescope and the SIRIUS camera to conduct our survey.
The IRSF (InfraRed Survey Facility) is a 1.4-m telescope
and the SIRIUS (Simultaneous 3-color InfraRed Imager for Unbiased Survey)
is a near-infrared camera which
can take images in three photometric bands ($JH\Ks$) simultaneously. 
The field of view is about $7.7\arcmin \times 7.7\arcmin$ with
the pixel scale of $0.45\arcsec$/pix.
Details of the instrument can be found in \citet{Nagashima-1999}
and \citet{Nagayama-2003}.

We observed three different lines of sight within the Galactic plane
($b=0\degr$): $l= +40\degr, +30\degr$ and $+20\degr$.
We hereby denote the three regions Lp40, Lp30 and Lp20, respectively.
At around each Galactic longitude, we observed nine fields-of-view
covering $20^{\prime}\times 20^{\prime}$.
Time-series observations were performed for approximately 45 times between
2007 and 2012.
For each set of monitorings, we took five exposures with 8~sec
integration in total.

To search for Cepheids, the methods in previous studies
\citep{Matsunaga-2009,Matsunaga-2013} are adopted.  
The limiting magnitudes are 15.5, 15.1 and 14.3~mag in $\JHK$
for the Lp20 region, and slightly deeper, by up to 0.3~mag,
in the other two regions
due probably to stellar blending caused by the higher stellar density
for Lp20.
The saturation limits are around 8.5, 8.5, 8.0~mag in $\JHK$.
We detected approximately 45000, 30000, 25000 stars in
the regions Lp20, Lp30 and Lp40, respectively,
although the numbers of stars detected in $J$ are roughly half of
those in the other two bands.
Among 300 candidates of variable star identified based on
their large standard deviations of time-series photometric results,
we have identified roughly 50 variables whose periods are
shorter than 60~days.

\subsection{Our targets}

In this paper, we report 3 classical Cepheids, one in the Lp20
and two in the Lp30 region, for which high-resolution spectra were 
obtained as we describe in Section~\ref{sec:spec}. 

\begin{figure}
\begin{center}
\includegraphics[clip,width=0.9\hsize]{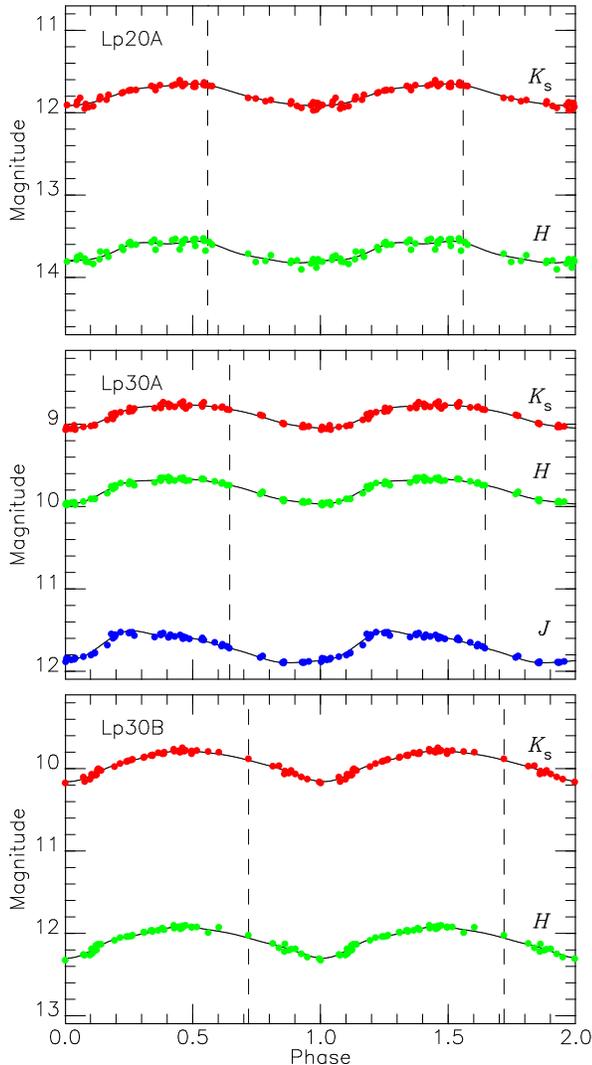}
\end{center}
\caption{
Light curves of our classical Cepheids.
Vertical dashed lines indicate the phases of
the IRCS spectroscopic observations (Section~\ref{sec:spec}).
\label{fig:LCs}}
\end{figure}

Table~\ref{tab:cat} lists their coordinates,
periods and mean magnitudes. The mean magnitudes were obtained
as intensity means of maximum and minimum from fourth-order Fourier fits
(see eq.~1 in \citealt{Matsunaga-2013}).
Figure~\ref{fig:LCs} plots their light curves together with the Fourier fits
and Table~\ref{tab:phot} lists time-series photometric data for
the three objects.
Lp20A and Lp30B are below the detection limit in $J$.
We also consider the mid-infrared magnitudes in
3.6 and 4.5~$\mu$m,
accessible at the VizieR server\footnote{
This research has made use of the VizieR catalog access tool, CDS,
Strasbourg, France. The original description of the VizieR service was
published in Ochsenbein {et~al.} (2010, A\&AS 143, 23).},
in the catalog of the GLIMPSE survey carried out with
the {\it Spitzer} space telescope
\citep{Churchwell-2009,IPAC-2008}.

\begin{deluxetable}{cccc}
\tablewidth{0pt}
\tablecaption{Catalog of new Cepheids\label{tab:cat}}
\tablehead{
 & \colhead{Lp20A} & \colhead{Lp30A} & \colhead{Lp30B}}
\startdata
RA~(J2000) & 18:28:11.47 & 18:45:28.56 & 18:45:48.01 \\
Dec~(J2000) & $-$11:37:32.4 & $-$02:44:40.5 & $-$02:27:30.1 \\
$l~(\degr)$ & 19.9541 & 29.8099 & 30.1015 \\
$b~(\degr)$ & $-$0.2073 & $+$0.0738 & $+$0.1325 \\
$P$~(days) & 10.79 & 12.72 & 42.70 \\
$\langle J\rangle$~(mag) & --- & 11.72 & --- \\
$\langle H\rangle$~(mag) & 13.70 & 9.83 & 12.13 \\
$\langle \Ks\rangle$~(mag) & 11.79 & 8.92 & 9.99 \\
$[3.6]$~(mag) & 10.45 & 8.16 & 8.65 \\
$[4.5]$~(mag) & 10.18 & 8.00 & 8.44\enddata
\tablecomments{
Listed for the three target Cepheids are
positions (RA and Dec in J2000 together with 
the Galactic coordinate $l, b$), period $P$,
mean magnitudes in $J$, $H$ and $\Ks$
from IRSF/SIRIUS, and
$[3.6]$ and $[4.5]$ magnitudes taken from
the GLIMPSE catalog (see text).
}
\end{deluxetable}

\begin{deluxetable}{crrrr}
\tablewidth{0pt}
\tablecaption{Photometric Data for Cehpeids\label{tab:phot}}
\tablehead{
\colhead{Object} & \colhead{MJD} & \colhead{$J$}
& \colhead{$H$} & \colhead{$\Ks$} \\ 
\colhead{} & \colhead{} &\colhead{(mag)} &\colhead{(mag)} &\colhead{(mag)}
}
\startdata
Lp20A & 54226.02957 & \multicolumn{1}{c}{---} & 13.54 & 11.68
 \\ 
Lp20A & 54229.06910 & \multicolumn{1}{c}{---} & 13.80 & 11.86
 \\ 
Lp20A & 54231.08360 & \multicolumn{1}{c}{---} & 13.78 & 11.87
 \\ 
Lp20A & 54231.18865 & \multicolumn{1}{c}{---} & 13.81 & 11.88
 \\ 
Lp20A & 54232.02653 & \multicolumn{1}{c}{---} & 13.73 & 11.82
 
\enddata
\tablecomments{
This is the first five lines of 162 $\JHK$ measurements
in total for three objects, Lp20A, Lp30A and Lp30B.
The entire table will only be available online.
}
\end{deluxetable}

We also searched for mid-infrared photometry for our Cepheids in
the ALLWISE catalog \citep{Cutri-2013}.
Lp20A is too faint in the WISE images and no magnitudes are
given in the catalog. Lp30A is relatively bright and
its $W1$ and $W2$ magnitudes are 
consistent with the $[3.6]$ and $[4.5]$, respectively,
within ${\sim}0.1$~mag in spite of the time variation and
the difference between the photometric systems\footnote{The effective
wavelengths of these mid-infrared bands are: 3.35 and 4.60~{$\mu$m} for $W1$ and $W2$ \citep{Wright-2010}, and 3.55 and 4.48~{$\mu$m} for $[3.6]$ and $[4.5]$ \citep{Hora-2008}.
}.
For Lp30B, the GLIMPSE image shows a few infrared sources
located close to each other around the target,
but the WISE catalog only gives one (blended) photometric measurement
in each band. The angular resolution of the WISE,
$6.1\arcsec$ in $W1$ \citep{Wright-2010},
is lower than that of the GLIMPSE survey from the {\it Spitzer},
$1.6\arcsec$ in $[3.6]$
\citep{Churchwell-2009}. Considering the limited benefit,
we decided to discard the WISE data in the following discussions.

\subsection{Distances and the effect of the extinction law}

The $H$- and $\Ks$-band mean magnitudes and the $[3.6]$
single-epoch magnitudes can be
used to determine the distances
to our newly discovered Cepheids, 
by adopting PLR and extinction coefficients in the respective bands. 
As discussed in \citet{Matsunaga-2016}, 
in the case of heavily reddened objects through the Galactic disk,
the largest source of error on the obtained distances will be the    
uncertainty on the adopted extinction law, which provides
the extinction coefficients. 
In order to quantify the impact of such uncertainty on
the derived distances, we devised a new technique,
presented in Inno {et~al.} (in prep). In summary, we
adopt the PLR of classical Cepheids
from \citet{Matsunaga-2013} for $H$ and $\Ks$ while the
$\log P$--$[3.6]$ relation is taken from \citet{Marengo-2010}:
\begin{eqnarray}
M_{H}=-3.256 (\log P -1.3) -6.562 \label{eq:PHR} \\
M_{\Ks}=-3.295 (\log P -1.3) -6.685 \label{eq:PKR} \\
M_{[3.6]}=-3.16 (\log P -1.0) -5.74 \label{eq:P3R} 
\end{eqnarray}
All of these relations were calibrated based on
trigonometric parallaxes
of Cepheids in the solar neighborhood \citep{Benedict-2007,vanLeeuwen-2007}.
Then, we use the extinction coefficients, $\AKEHK$ or $\AKEKT$,
the observed magnitudes and the PLR 
in two bands, either $H$ and $\Ks$ or $\Ks$ and $[3.6]$,
to estimate the distance modulus $\mu_0$ and foreground extinction $\AK$.
Note that the $[4.5]$ magnitudes are not used  
because the $\log P$--$[4.5]$ relation shows a metallicity dependence
related to the  presence of the CO absorption at this wavelength
\citep{Hackwell-1974,Majaess-2013,Scowcroft-2016}.
In contrast, the metallicity effect on the PLR in the range of
$H$, $\Ks$ and $[3.6]$ is minimal if any \citep{Bono-2010,Scowcroft-2016}.
While the characterization of the infrared PLR to a very high precision is important and still on progress, e.g.~in a cosmological context \citep{Riess-2016}, the accuracy of our analysis is limited by the reddening correction as we see below.
Finally, a comparison of the multiple estimates using both near- and
mid-infrared magnitudes and using two different extinctions laws
allows us to evaluate such errors in a robust manner.

We consider two $\AKEHK$ values:
1.44 from \citet{Nishiyama-2006} and 1.83 from \citet{Cardelli-1989}.
These values are at around two extremes of the extinction law
for the near-infrared range among previous investigations
\citep[see e.g.\  table 1 of][]{Nishiyama-2006}.
For $\AKEKT$, we adopt 2.01 from \citet{Nishiyama-2009}
and 1.76 predicted by the power law with the index of
1.61 \citep{Cardelli-1989}.
3.6~$\mu$m is slightly outside the near-infrared regime
considered by \citet[0.9--3.33~$\mu$m]{Cardelli-1989},
but we only use these $\AKEKT$ values for
comparing the effect of two substantially different extinction laws.

It should be noted that the results of \citet{Nishiyama-2006,Nishiyama-2009}
were obtained with giants in the Galactic bulge while
those of \citet{Cardelli-1989} were based on stars
in the solar neighborhood.
There can be a spatial variation
of the reddening law in a rather complicated manner
\citep[see e.g.][and reference therein]{Nataf-2016b}.
Unfortunately, the extinction law toward
the regions of our interest, $l \approx 20-30\degr$ 
at several kpc, is not well established.
The anonymous referee pointed out that most of the recent results \citep[e.g.][]{AlonsoGarcia-2015,Majaess-2016} on the extinction law in the infrared regime are different from that of \citet{Cardelli-1989} \citep[see also reviews and discussions by][]{Nataf-2016a,Damineli-2016}. Some works, however, suggest that the extinction laws show variations depending on sight-lines and/or the amount of extinction \citep{Nishiyama-2006,Fitzpatrick-2009,Gosling-2009}. It is hard to give a robust range of its uncertainty, and here we use the two extreme laws, \citet{Nishiyama-2006} and \citet{Cardelli-1989}, to give likely maximum uncertainties.

\begin{deluxetable}{cccc}
\tablewidth{0pt}
\tablecaption{Estimates of distance modulus $\mu_0$ (and extinction $\AK$)\label{tab:DMAK}}
\tablehead{
Object  & Lp20A & Lp30A & Lp30B}
\startdata
\multicolumn{4}{c}{{\it Assumption: Classical Cepheid}} \\
\multicolumn{1}{l}{(1)~$H, \Ks$ \& N06} & 15.00~(2.59) & 13.81~(1.15) & 14.86~(2.89) \\
\multicolumn{1}{l}{(2)~$H, \Ks$ \& C89} & 14.30~(3.29) & 13.50~(1.46) & 14.08~(3.67) \\
\multicolumn{1}{l}{(3)~$\Ks, [3.6]$ \& N09} & 14.98~(2.61) & 13.50~(1.46) & 15.00~(2.76) \\
\multicolumn{1}{l}{(4)~$\Ks, [3.6]$ \& C89} & 15.31~(2.29) & 13.68~(1.28) & 15.34~(2.42) \\
\multicolumn{1}{l}{(5)~Combined} & $14.90 \pm 0.25$ & $13.62 \pm 0.09$ & $14.82 \pm 0.31$ \\
 & ($2.70 \mp 0.24$) & ($1.34 \mp 0.09$) & ($2.94 \mp 0.31$) \\
\hline
\multicolumn{4}{c}{{\it Assumption: Type II Cepheid}} \\
\multicolumn{1}{l}{(6)~$H, \Ks$ \& N06} & 12.58~(2.63) & 11.31~(1.19) & 11.78~(2.99) \\
\multicolumn{1}{l}{(7)~$H, \Ks$ \& C89} & 11.87~(3.34) & 10.98~(1.52) & 10.97~(3.80)\enddata
\tablecomments{
For each Cepheid given are four estimates (1--4) of
distance modulus $\mu_0$ and extinction $\AK$
with different combinations of two photometric magnitudes,
either $H$ and $\Ks$ or $\Ks$ and $[3.6]$,
and the extinction law, C89 \citep{Cardelli-1989}, N06
\citep{Nishiyama-2006} or N09 \citep{Nishiyama-2009}.
The raw (5) gives the means and standard errors,
of the four estimates, which are used in the discussions 
(double signs correspond to one another for each object).
The rows (6) and (7) give the estimates under the assumption of type II Cepheids which are used for Figure~\ref{fig:DAK}.
} 
\end{deluxetable}

\begin{figure}
\begin{center}
\includegraphics[clip,width=0.9\hsize]{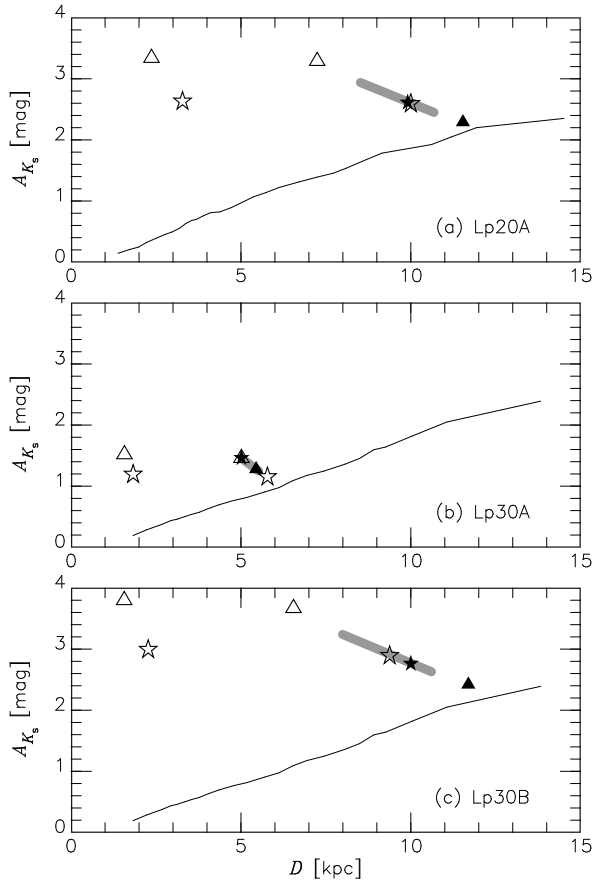}
\end{center}
\caption{
The three panels present the comparison of
the current estimates of extinction and distance
for the three targets (Lp20A, top; Lp30A, middle, Lp30B, bottom)
with the three-dimensional extinction map (solid curve) provided by
\citet{Marshall-2006}.
Open and filled symbols indicate estimates based on the $H, \Ks$ data
and those based on the $\Ks, [3.6]$ data, respectively.
Triangles and star symbols indicate the estimates by adopting
the extinction laws of \citet{Cardelli-1989} and
the law of \citet{Nishiyama-2006,Nishiyama-2009}
respectively.
The symbols located at $D>4$~kpc are obtained 
by assuming that the objects are
classical Cepheids, while the others by assuming that they are type II Cepheids.
The thick gray lines mark the ``combined'' estimates
with the assumption of classical Cepheids listed in
Table~\ref{tab:DMAK}.
\label{fig:DAK}}
\end{figure}

Table~\ref{tab:DMAK} lists the estimates of $(\mu_0, \AK)$
obtained with different combinations of photometric bands and
different extinction laws (raws 1--4).
Also given are the combined values (row 5) which are
simple means (and standard errors) of the four estimates,
and we use them in the discussions in Section~\ref{sec:Discussion}.
The error budgets of these combined values are dominated by
the uncertainty of the extinction law.
The different $\AKEHK$ values from \citet{Nishiyama-2006} and
\citet{Cardelli-1989} lead to significantly different
$\AK$ and $\mu _0$ values, {e.g.}~${\sim}0.8$~mag for
Lp30B~(Table~\ref{tab:DMAK}).
Typical peak-to-valley amplitudes $\Delta [3.6]$ of 0.2~mag in
the $[3.6]$ band \citep{Monson-2012} result in errors of $\pm 0.1$~mag
because we used single-epoch magnitudes of $[3.6]$, but
those errors are smaller than the uncertainty introduced
by the extinction law for Lp20A and Lp30B.
The effect of $\Delta [3.6]$ is comparable to the uncertainty due to
the extinction law in case of Lp30A whose reddening is not so large.
Figure~\ref{fig:DAK} plots the extinction $\AK$ against the
distance $D$ based on the estimates with
the different datasets and the extinction laws (Nishiyama versus Cardelli).
The large scatters of the estimates clearly illustrate
the impact of the extinction law.

Table~\ref{tab:DMAK} and Figure~\ref{fig:DAK} include estimates
under the assumption that our targets are type II Cepheids.
Such an assumption, however, can be rejected even if the choice of
the extinction law has a large impact on the distance estimates (raws 6--7).
Because of similar colors but distinctly different absolute magnitudes
for classical Cepheids and type II Cepheids
\footnote{
We used the PLR of type II Cepheids taken from \citet{Matsunaga-2011a} as described in the Supplementary Information of \citet{Matsunaga-2011b}.
}
at a given period, assuming the two different types of Cepheids would lead to
similar reddenings but totally different distances (Table~\ref{tab:DMAK}).
This allows us to determine the types of Cepheids
\citep{Matsunaga-2011b,Matsunaga-2014,Matsunaga-2016}.
Lp20A and Lp30B are much redder than intrinsic colors of Cepheids
and obviously affected by very large extinctions,
2.5--3~mag in $\AK$ or even larger.
The reddening of Lp30A is moderate, slightly larger than $\AK = 1$~mag.
Their distances would be estimated to be within 4~kpc
if they were type II Cepheids (Figure~\ref{fig:DAK}),
but at these short distances such large extinctions are far from expected.
The $(D, \AK)$ values obtained by assuming that these stars
are classical Cepheids are closer to the three-dimensional extinction map
provided by \citet{Marshall-2006}, 
who gives $\AK = 1.8$~mag at 10~kpc toward the redder two,
Lp20A and Lp30B, and 0.9~mag at 5.5~kpc toward Lp30A.
With the short distances obtained with the PLR of type II Cepheids,
$D \la 3$~kpc, their extinction map gives $\AK\sim 0.5$~mag,
much smaller than the values in Table~\ref{tab:DMAK}.

Further investigations of spectra may be useful in giving
further supports on the classification as classical Cepheids.
Type II Cepheids are, for example, relatively metal poor
compared to classical Cepheids in general. While some type II Cepheids
are as metal-rich as the solar \citep{Maas-2007},
classical Cepheids in the inner disk are expected to be more metal-rich
\citep[e.g.][]{Genovali-2014}.

\section{Spectroscopic observations}
\label{sec:spec}

\begin{figure*}
\begin{center}
\includegraphics[clip,width=0.75\hsize]{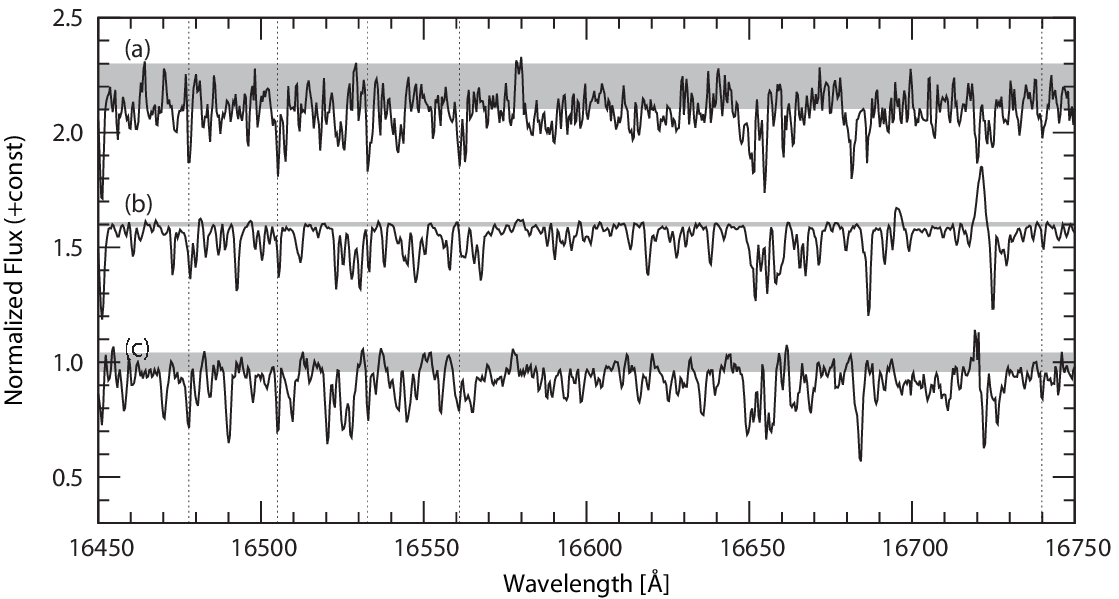}
\end{center}
\caption{
A part of Subaru/IRCS spectra in the $H$ band for our Cepheids,
(a)~Lp20A, (b)~Lp30A and (c)~Lp30B.
The grey horizontal strip for each spectrum indicates the continuum with
the width of $\pm 1\sigma$ according to the S/N in Table~\ref{tab:spec}.
Vertical dotted lines indicate telluric lines.
With an eye check on raw data, we found that
the seemingly emission line at around 16730{\AA}  for Lp30A
and other less prominent positive features are not real.
\label{fig:spec}}
\end{figure*}

We observed the three Cepheids with the Infrared Camera and
Spectrograph (IRCS) equipped with the adoptive optics system, AO188,
and attached to the Subaru 8.2~m telescope
\citep{Kobayashi-2000,Hayano-2010}. This instrument allowed us to 
obtain high-resolution ($\lambda/\Delta\lambda = 20,000$)
$H$-band spectra. 
The observation log is given in Table~\ref{tab:spec}.
The signal-to-noise ratios, S/N, are significantly different among
the three targets as listed in the table
(also see the spectra in Figure~\ref{fig:spec}).

\begin{deluxetable}{cccc}
\tablewidth{0pt}
\tablecaption{Log of Spectroscopic Observations and Measured Radial Velocities of the Cepheids\label{tab:spec}}
\tablehead{
 & \colhead{Lp20A} & \colhead{Lp30A} & \colhead{Lp30B}}
\startdata
Date~(UTC) & 2012-07-28 & 2012-07-27 & 2012-07-27 \\
Time\tablenotemark{a}~(UTC) & 11:20 & 10:45 & 11:40 \\
MJD\tablenotemark{b} & 56136.47 & 56135.45 & 56135.48 \\
Phase & 0.56 & 0.64 & 0.72 \\
Integrations\tablenotemark{c} & 300$^{\rm s} \times 16$ & 120$^{\rm s} \times 8$ & 300$^{\rm s} \times 12$ \\
S/N & 10 & 100 & 25 \\
$\sigma_{V}~(\kms)$ & 1.1 & 0.7 & 1.3 \\
$\Delta V~(\kms)$ & -3 & 3 & 15 \\
\hline
$V_\mathrm{helio}~(\kms)$ & 7 & 93 & 32 \\
$\VLSR~(\kms)$ & 22 & 109 & 49\enddata
\tablenotetext{a}{Coordinated Universal Time at around the middle of the observation}
\tablenotetext{b}{Modified Julian Date calculated from the given Time}
\tablenotetext{c}{Duration of each integration ($t_1$) and the number of integrations $N$}
\tablecomments{
The heliocentric velocities ($V_\mathrm{helio}$) and
velocities relative to the LSR ($V_\mathrm{LSR}$)
were calculated from measurements over five echelle orders
with the standard error $\sigma_{V}$, and are
after the correction of pulsational velocity $\Delta V$ subtracted.
} 
\end{deluxetable}

The data reduction and radial velocity measurements
were carried out following the steps described in \citet{Matsunaga-2015}.
In short, a spectrum with $N$ raw spectra combined after the normalization
was prepared for each object (Figure~\ref{fig:spec}), and
cross-correlated with synthetic spectra
which include both stellar absorption lines and telluric lines
to search for the velocity that best explains the shift between
the stellar and telluric lines.
We estimated the velocity with five echelle orders
and for each target these orders gave us self-consistent values
leading to a small standard error of mean,
$\sigma _V$ listed in Table~\ref{tab:spec},
even for Lp20A whose spectrum has a rather low S/N.

It is necessary to make corrections of the pulsational effect
to investigate the kinematics of the Cepheids in the disk.
This was done following the same method described in
\citet{Matsunaga-2015} but using templates of velocity and light curve
constructed for Cepheids in each period range of a target.
The detail of the templates will be given in Inno {et~al.} (in prep).
Table~\ref{tab:spec} lists the estimated offsets $\Delta V$
in velocity to be added for the correction.
Thus corrected radial velocities were transformed into
the heliocentric values $\Vhelio$ and 
the velocities relative to the local standard of rest, $\VLSR$,
assuming the standard solar motion \citep{Crovisier-1978,Reid-2009}. 
The errors of the $\VLSR$ are dominated by
the uncertainty in the correction of the pulsational effect,
and are estimated to be 10~$\kms$.

\section{Discussion}
\label{sec:Discussion}

It is known that classical Cepheids show a narrow distribution
with a scale height of about ${\sim}100$~pc
around the Galactic plane, 
and our Cepheids are found within $0.25\degr$ in Galactic latitude
due also to the limited survey area. We compare their distribution and
kinematics with those of the HMSFRs
with parallaxes measured by \citet{Reid-2014}. 
The latter group is also located in a similarly  
small range around the Galactic midplane, within $0.5\degr$ except
a few nearby objects.

\begin{figure*}
\begin{center}
\includegraphics[clip,width=0.57\hsize]{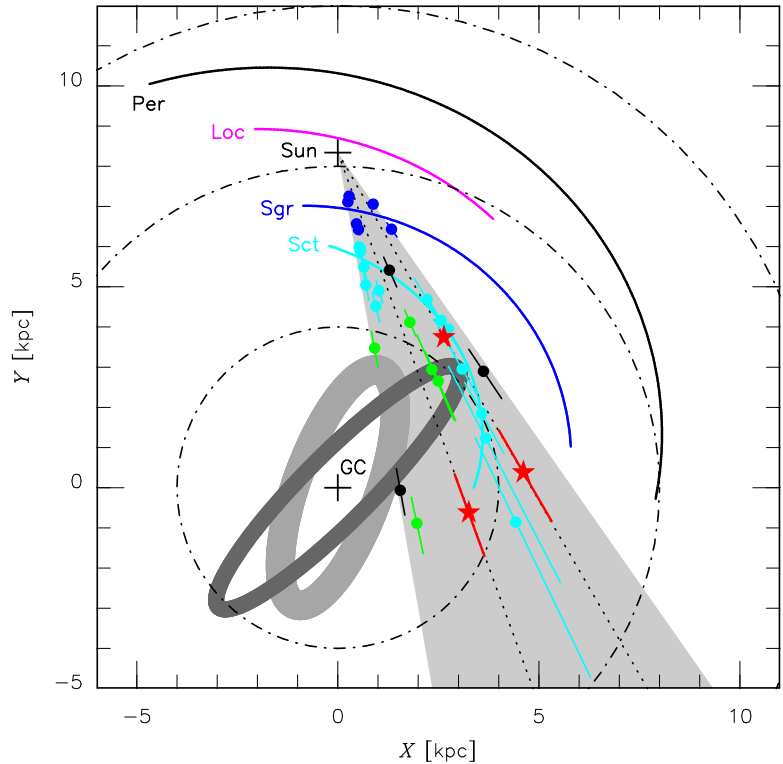}
\end{center}
\caption{
Distribution of our three Cepheids are indicated by red star symbols. 
The locations of high-mass star forming regions, between $10\degr$
and $35\degr$ in $l$, with trigonometric parallaxes measured are taken from
\citet{Reid-2014} and indicated by circles with different colors
according to spiral arms or other components to which the objects are assigned:
inner Galaxy sources, green; Scutum arm, cyan; Sagittarius arm, blue;
others, black. The short and long Galactic bars are also adopted from
\citet{Reid-2014} and indicated by grey ellipses.
Circles are drawn to mark the Galactocentric distances
of 4, 8 and 12~kpc.
\label{fig:Map}}
\end{figure*}

Figure~\ref{fig:Map} plots the distributions of the Cepheids
and the HMSFRs in the range of $l=10-35\degr$ 
projected onto the Galactic plane. We draw, as a guide,
the spiral arms and Galactic bars which are
also adopted from \citet{Reid-2014}
except the Outer arm which is not relevant to this work.
Lp30A is clearly located along the Scutum arm;
its distance error is relatively small, at least comparable to
the parallax-based distances to nearby HMSFRs
and has little effect on the proximity of Lp30A
in the tangential direction of the Scutum arm.
Although the other two Cepheids seem located near the inner part of
the Scutum arm, the larger errors of their distances 
make it hard to draw a firm conclusion on their locations. Spiral arms 
in the Galactic disk have not been established at around 10~kpc from the Sun
and beyond even with gas and star-forming components like \ion{H}{2} regions.

\begin{figure}
\begin{center}
\includegraphics[clip,width=0.88\hsize]{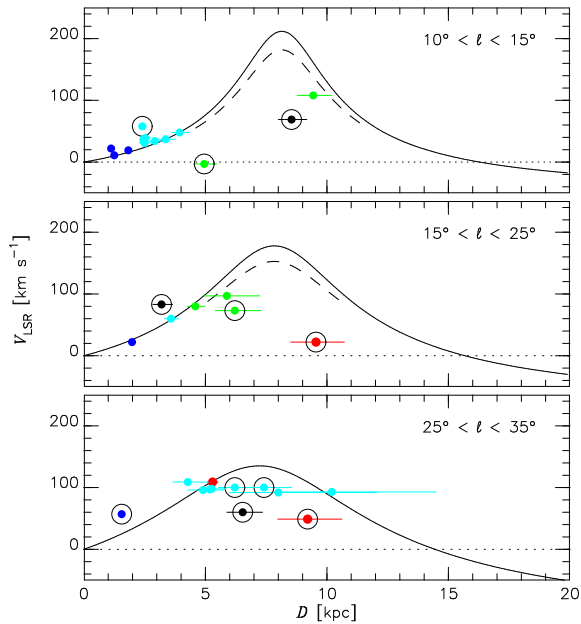}
\end{center}
\caption{
$\VLSR$ of the Cepheids and those of high-mass star forming regions 
from \citet{Reid-2014} are plotted against distance.
The filled curve indicates the velocity predicted by
the flat rotation with $\Theta_0 = 240~\kms$, while the dashed curve
adopts the velocity smaller by 15~\% within 5~kpc of the Galactic center.
Symbols are the same as those in Figure~\ref{fig:Map},
but encircled are objects whose $\VLSR$ are significantly different from
the filled curve (see text, Table~\ref{tab:drift}
and Figure~\ref{fig:Vlag3}).
\label{fig:DV}}
\end{figure}

Figure~\ref{fig:DV} plots $\VLSR$ against distance from the Sun
for the Cepheids and the HMSFRs
grouped into three ranges of Galactic longitude.
Assuming the circular rotation of the disk, we expect
\begin{equation}
\VLSR = \left( \frac{\Theta}{R} -\frac{\Theta_0}{R_0} \right) R_0 \sin l
\end{equation}
where $\Theta$ and $R$ are the circular orbital speed and
the distance from the Galactic center ($\Theta _0$
and $R_0$ for the solar position), and the distance from the Sun, $D$,
can be translated to the Galactocentric distance as
\begin{equation}
R=\sqrt{D^2+R_0^2-2DR_0\cos l}.
\end{equation}
The solid curve in Figure~\ref{fig:DV} shows thus calculated
$\VLSR$ in each direction where the solar values are adopted from
\citet{Reid-2014}: $\Theta _0 = 240 (\pm 8)~\kms$ and
$R_0 = 8.34 (\pm 0.16)$~kpc.
It has been suggested that the Galactic rotation is slower
in the innermost part of the disk (see, for example,
figure~4 in \citealt{Reid-2014}). The dashed curve in Figure~\ref{fig:DV}
indicates how the expected $\VLSR$ changes if $\Theta$ gets smaller
by 15~\%, i.e.~36~$\kms$, within 5~kpc of the Galactic center.

\begin{figure}
\begin{center}
\includegraphics[clip,width=0.88\hsize]{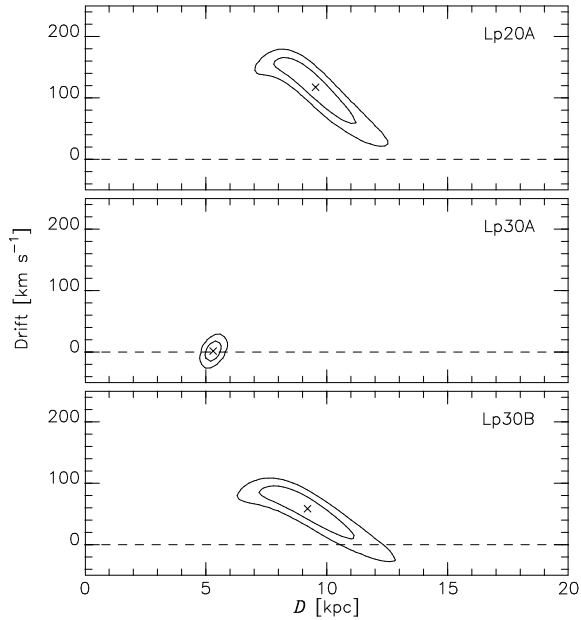}
\end{center}
\caption{
Results of the Monte-Carlo simulations performed to compare
the $\VLSR$ of Cepheids with the Galactic rotation (see text for detail).
The contours indicate the areas enclosing 68.26~\% ($1~\sigma$) and
95.44~\% ($2~\sigma$) of $10^6$ trials for each object.
The drift is positive (or negative) when 
the object is moving away from us slower (or faster) than expected
from the Galactic rotation, while the zero drift indicated by
horizontal line indicates that the object follows the Galactic rotation.
\label{fig:Vlag3}}
\end{figure}

Some objects show offsets from the predicted curve of $\VLSR$
in Figure~\ref{fig:DV}.
A method of Monte-Carlo simulation is used to judge
whether $\VLSR$ is different from the prediction of
the Galactic rotation and to estimate how large the drift 
from the rotation curve is, in a statistical manner, as follows.
We consider the distance and $\VLSR$ as well as their errors
for each object, and generate the trial values, $d$ and $v$,
with simulated Gaussian errors added in each run.
Then, a $\VLSR$ at the distance $d$ is predicted based on
the Galactic rotation model but with simulated values of
$R_0$ and $(\Theta_0 + V_\odot)/R_0$ taking their errors into account.
The solar motion in the direction of the Galactic rotation, $V_\odot$,
is fixed as the standard value of 15.4~$\kms$
which is used for the translation from $\Vhelio$ to $\VLSR$ in this work.
The $(\Theta_0 + V_\odot)/R_0$ parameter, $30.75\pm 0.43~\kms~{\rm kpc}^{-1}$
\citep{Reid-2014},
is used as an independent variable instead of $\Theta_0$ itself
because the measurements of $R_0$ and $\Theta_0$ may be 
correlated \citep{Reid-2009,Reid-2014}.
Thus simulated $\VLSR$ value, $v^\prime$, at the distance of $d$
is compared with $v$, and the difference $v^\prime -v$ is taken as
a simulated value of the drift. The drift has a positive value if
the object is moving away from us slower than expected with
the Galactic rotation.
We repeat such a run $10^6$ times and check how
the simulated $(d, v^\prime -v)$ are distributed.
Figure~\ref{fig:Vlag3} plots such distributions for the three Cepheids
as contours. Table~\ref{tab:drift} lists the median values of
the simulated drifts, $v-v^\prime$ and the ranges of $\pm 1\sigma$.
We also calculated how much a deviation from the zero drift
is significant by counting the simulation runs which gave zero or negative
(or instead positive) drift
if the object has a positive (negative) median drift.
For example, less than ${\sim}0.1~\%$ of the $10^6$ runs returned
drift of zero or smaller for Lp20A for which the median drift was
estimated $117~\kms$, which rules out the null hypothesis that
this object is moving as expected by the Galactic rotation (or even faster)
by the 3.1~$\sigma$ significance.
Lp30B also has a significant positive drift.
In contrast, Lp30A has a negligible drift
and follows the Galactic rotation.
Together with the comparisons to the Galactic rotation,
the Monte-Carlo simulations provide us with estimates of the Galactocentric
distances of Cepheids taking various errors into account;
we obtained $\RGC=3.3^{+0.7}_{-0.4}$~kpc, $4.6\pm 0.16$~kpc
and $4.6^{+0.8}_{-0.4}$~kpc for Lp20A, Lp30A and Lp30B, respectively.
Our three classical Cepheids are among those nearest to the Galactic center
\citep{Dekany-2015b,Matsunaga-2016}
except the ones found in the nuclear stellar disk
\citep{Matsunaga-2011b,Matsunaga-2015}.

We performed the same kind of Monte-Carlo simulations
for HMSFRs from \citet{Reid-2014}
within the range of 10--$35\degr$ (Table~\ref{tab:drift}).
As pointed out by \citet{Reid-2014}, some HMSFRs show
significant drifts, or peculiar motions, from the Galactic rotation.
While the position and kinematics of Lp30A
are perfectly consistent with
those of surrounding HMSFRs in the Scutum arm,
we found that large drifts for the other two Cepheids.
For Lp20A, there are no comparable young tracers 
due to the absence of HMSFRs with maser parallaxes measured in its direction.
At its Galactocentric distance, ${\sim}3.5$~kpc, the Galactic rotation may
be deviated from the flat rotation. There are two HMSFRs
which are located even nearer to the Galactic center:
G010.47$+$00.02 is assigned to Connecting arm and G012.02$-$00.03 to 
Far 3-kpc arm \citep{Sanna-2014}. 
These arms are considered to be related to the Galactic bar(s),
in particular the long thin bar, and their extensions in
Galactic longitude are between $-12\degr$ and $+13\degr$
\citep{Dame-2008,Sanna-2014}.
Lp20A is located away from these inner structures.
Lp30B has an even larger Galactocentric distance, ${\sim}5$~kpc;
a few SFRs relatively close to it seem to show
$\VLSR$ more-or-less consistent with the Galactic rotation but
the measurements of their parallaxes leave larger uncertainties
than that for Lp30B. 
The two Cepheids with large drifts are located at around
the interface between the disk and the bulge where the non-axisymmetric
gravitational perturbations may have strong effects on gas and stellar orbits.
The very limited statistics of Cepheids in such
an inner region of the Galaxy and the lack of their proper motions
prevent us from drawing conclusions on the general features
of stellar kinematics. Tracers like Cepheids in such regions, nevertheless,
would give important constraints on the stellar kinematics in the inner Galaxy,
and the drift considered in this work would be a useful indicator
for such studies.

\begin{deluxetable}{cccc}
\tablewidth{0pt}
\tablecaption{Drifts of Cepheids and high-mass star forming regions from the Galactic rotation\label{tab:drift}}
\tablehead{
\colhead{Object} & \colhead{Spiral} & \colhead{Drift} & \colhead{Deviation}\\  
\colhead{} & \colhead{Arm} & \colhead{($\kms$)} &\colhead{} 
}
\startdata
Lp20A & --- & $117^{+31}_{~-43}$ & slow ~($3.1~\sigma$) \\ 
Lp30A & Sct & $2^{+11}_{~-11}$ &  \\ 
Lp30B & --- & $59^{+23}_{~-35}$ & slow ~($1.6~\sigma$) \\ 
\hline
G010.47$+$00.02 & Con & $139^{+14}_{~-34}$ & slow ~($3.5~\sigma$) \\ 
G010.62$-$00.38 & 3-k & $69^{+18}_{~-13}$ & slow ~($>5.0~\sigma$) \\ 
G011.49$-$01.48 & Sgr & $-2^{+3}_{~-3}$ &  \\ 
G011.91$-$00.61 & Sct & $-1^{+9}_{~-7}$ &  \\ 
G012.02$-$00.03 & 3-k & $53^{+39}_{~-45}$ &  \\ 
G012.68$-$00.18 & Sct & $-35^{+10}_{~-10}$ & fast ~($3.4~\sigma$) \\ 
G012.80$-$00.20 & Sct & $-3^{+8}_{~-7}$ &  \\ 
G012.88$+$00.48 & Sct & $-6^{+8}_{~-8}$ &  \\ 
G012.90$-$00.24 & Sct & $-12^{+10}_{~-10}$ &  \\ 
G012.90$-$00.26 & Sct & $-14^{+10}_{~-10}$ &  \\ 
G013.87$+$00.28 & Sct & $5^{+14}_{~-13}$ &  \\ 
G014.33$-$00.64 & Sgr & $-12^{+5}_{~-5}$ &  \\ 
G014.63$-$00.57 & Sgr & $-1^{+5}_{~-5}$ &  \\ 
G015.03$-$00.67 & Sgr & $-1^{+4}_{~-3}$ &  \\ 
G016.58$-$00.05 & Sct & $-8^{+9}_{~-8}$ &  \\ 
G023.00$-$00.41 & 4-k & $11^{+12}_{~-10}$ &  \\ 
G023.44$-$00.18 & 4-k & $32^{+26}_{~-28}$ &  \\ 
G023.65$-$00.12 & --- & $-29^{+12}_{~-9}$ & fast ~($2.0~\sigma$) \\ 
G023.70$-$00.19 & 4-k & $66^{+18}_{~-23}$ & slow ~($3.0~\sigma$) \\ 
G025.70$+$00.04 & Sct & $6^{+48}_{~-93}$ &  \\ 
G027.36$-$00.16 & Sct & $33^{+19}_{~-78}$ &  \\ 
G028.86$+$00.06 & Sct & $34^{+12}_{~-15}$ & slow ~($1.7~\sigma$) \\ 
G029.86$-$00.04 & Sct & $27^{+8}_{~-12}$ & slow ~($2.1~\sigma$) \\ 
G029.95$-$00.01 & Sct & $12^{+12}_{~-12}$ &  \\ 
G031.28$+$00.06 & Sct & $-24^{+20}_{~-15}$ &  \\ 
G031.58$+$00.07 & Sct & $4^{+17}_{~-15}$ &  \\ 
G032.04$+$00.05 & Sct & $8^{+7}_{~-7}$ &  \\ 
G033.64$-$00.22 & --- & $57^{+5}_{~-6}$ & slow ~($3.5~\sigma$) \\ 
G034.39$+$00.22 & Sgr & $-30^{+6}_{~-5}$ & fast ~($>5.0~\sigma$) \\ 
G035.02$+$00.34 & Sgr & $-10^{+7}_{~-7}$ &  
\enddata
\tablecomments{
High-mass star forming regions located between $10\degr$ and $35\degr$ in $l$ are taken from \citet{Reid-2014}.
The assigned spiral arms for the star forming regions are taken from \citet{Reid-2014}, while the spiral arms for the Cepheids are based on this work (see text). 
The deviations of the Galactic rotation are indicated if the significance is larger than 1.5~$\sigma$} 
\end{deluxetable}

\section{Summary}

The three classical Cepheids we've reported here
are located at 3--5~kpc from the Galactic center.
Although the uncertainty in the extinction law leaves
large errors in distance (up to 15~\%), we found that the radial velocities
of two of the Cepheids are significantly different from those expected
from the Galactic rotation, possibly being influenced by the Galactic bar,
based on a method utilizing Monte-Carlo simulations.
Only a small number of Cepheids have been identified so far 
in such an inner part of the Galactic disk
\citep{Dekany-2015b,Matsunaga-2016},
and thus the new objects are interesting
for follow-up investigations. For example, their abundances will further
provide an insight into chemical evolution in the inner Galaxy. In the future,
significantly more Cepheids are expected to be found
in near-infrared variability surveys like VISTA Variables via Lactea 
\citep{Minniti-2010,Dekany-2015a,Dekany-2015b}.
The uncertainty in the extinction law, however, introduces large errors
in distances to the Cepheids deeply obscured in the Galactic disk.
It is necessary to determine
the extinction law in each direction of the disk in order to
construct an accurate map of Cepheids and other objects
in such obscured regions.

\acknowledgments

We acknowledge the anonymous referee for helpful comments to improve
the manuscript. We are grateful to the Subaru astronomer, Miki Ishii, 
who supported our observations in the 2012 July run.
This work has been supported by Grants-in-Aid for Scientific Research
(No.~22840008, 23684005, and 26287028)
from the Japan Society for the Promotion of Science (JSPS).
NM and LI acknowledge support from Sonderforschungsbereich SFB~881
``The Milky Way System'' (subproject A3) of the German Research Foundation (DFG).

{\it Facilities:} \facility{IRSF}, \facility{Subaru}

\end{document}